%% file: sample-edbt2025.tex
\documentclass[sigconf,edbt]{acmart-edbt2025}

\def\BibTeX{{\rm B\kern-.05em{\sc i\kern-.025em b}\kern-.08em
    T\kern-.1667em\lower.7ex\hbox{E}\kern-.125emX}}

\usepackage{booktabs} 
\usepackage{algorithm}
\usepackage{algorithmic}
\usepackage{graphicx}
\usepackage{textcomp}
\usepackage{multirow}
\usepackage{hyperref} 
\usepackage{enumitem} 
\usepackage{makecell} 
\usepackage{tabularx}
\usepackage{tcolorbox}
\usepackage{wrapfig}  
\usepackage[T1]{fontenc}

\setcopyright{rightsretained}

\acmDOI{}

\acmISBN{978-3-89318-099-8}

\acmConference[EDBT 2025]{28th International Conference on Extending Database Technology (EDBT)}{25th March-28th March, 2025}{Barcelona, Spain}
\acmYear{2025}

\begin{document}
\title{Automated Data Quality Validation in an End-to-End GNN Framework}

\newcommand{\tp}[1]{{\color{red} {\bf ??? #1 ???}}\normalcolor}
\newcommand{\qw}[1]{{\color{red} {\bf (qitong: #1)}}\normalcolor}
\newcommand{\sijie}[1]{{\color{blue} {\bf (sijie: #1)   }}\normalcolor}
\newcommand{\sj}[1]{{\color{red} {#1}}\normalcolor}
\newcommand{\soror}[1]{{\color{violet} {\bf (soror: #1)   }}\normalcolor}

\author{Sijie Dong}
\affiliation{%
  \institution{Université Paris Cité}
  \streetaddress{}
  \city{Paris}
  \country{France}
}
\email{sijie.dong@etu.u-paris.fr}

\author{Soror Sahri}
\affiliation{%
  \institution{Université Paris Cité}
  \city{Paris}
  \country{France}
}
\email{soror.sahri@parisdescartes.fr}

\author{Themis Palpanas}
\affiliation{%
  \institution{Université Paris Cité}
  \city{Paris}
  \country{France}
}
\email{themis@mi.parisdescartes.fr}

\author{Qitong Wang}
\affiliation{%
  \institution{Harvard University}
  \streetaddress{}
  \city{Boston}
  \country{USA}
}
\email{qitong@seas.harvard.edu}

\renewcommand{\shortauthors}{}

\begin{abstract}
Ensuring data quality is crucial in modern data ecosystems, especially for training or testing datasets in machine learning. 
Existing validation approaches rely on computing data quality metrics and/or using expert-defined constraints. 
Although there are automated constraint generation methods, they are often incomplete and may be too strict or too soft, causing false positives or missed errors, thus requiring expert adjustment. 
These methods may also fail to detect subtle data inconsistencies hidden by complex interdependencies within the data.
In this paper, we propose DQuag, an end-to-end data quality validation and repair framework based on an improved Graph Neural Network (GNN) and multi-task learning. The proposed method incorporates a dual-decoder design: one for data quality validation and the other for data repair.
Our approach captures complex feature relationships within tabular datasets using a multi-layer GNN architecture 
to automatically detect explicit and hidden data errors. 
Unlike previous methods, our model does not require manual input for constraint generation and learns the underlying feature dependencies, enabling it to identify complex hidden errors that traditional systems often miss. 
Moreover, it can recommend repair values, improving overall data quality. 
Experimental results validate the effectiveness of our approach in identifying and resolving data quality issues. The paper appeared in EDBT 2025.
\end{abstract}

%
%



\maketitle

\input{Sections/introduction}

\input{Sections/relatedwork}
\input{Sections/systems}
\input{Sections/evaluation}

\input{Sections/conclusion}

\begin{acks}
Work partially funded by EU Horizon projects AI4Europe (101070000),
TwinODIS (101160009), ARMADA (101168951), DataGEMS (101188416) and
RECITALS (101168490).
\end{acks}
\bibliographystyle{ACM-Reference-Format}
\bibliography{sample}

%
% ...

\end{document}

%% file: Sections/introduction.tex
\section{Introduction}
In the era of artificial intelligence and large models, ensuring data quality is crucial. High-quality data is essential for reliable decision-making, particularly in machine learning, where data quality directly impacts model performance. This paper addresses data quality validation, focusing on innovative approaches that overcome the limitations of existing methods.

\noindent\textbf{Background:}  
Quality validation is important to ensure that data meets specific requirements. Its main steps start
with data profiling~\cite{abedjan2017data}, generating metadata to describe quality like missing values and data dependencies~\cite{maydan1991efficient,naumann2014data}. 
This metadata forms the basis for assessing data quality against established metrics. 
However, this often relies on expert-defined constraints, which are resource-intensive and time-consuming to develop, maintain, and adjust.

Automated constraint generation~\cite{schelter2018automating,caveness2020tensorflow} attempts to learn from dataset characteristics, but these constraints are often incomplete or either too strict or too soft.
Strict constraints can lead to false positives, such as flagging acceptable minor data entry variations as errors. While soft 
constraints may miss critical issues, such as overlooking small discrepancies in numerical data. They need expert intervention for fine-tuning, which limits practical effectiveness.

Furthermore, these automated constraint generation methods often fail to detect hidden 
relationships and dependencies within the data, which we refer to as hidden errors. 
Indeed, data quality involves not only individual data points but also the complex interrelationships among them. 

\begin{figure}[tb]
\centering
\includegraphics[width=0.95\linewidth]{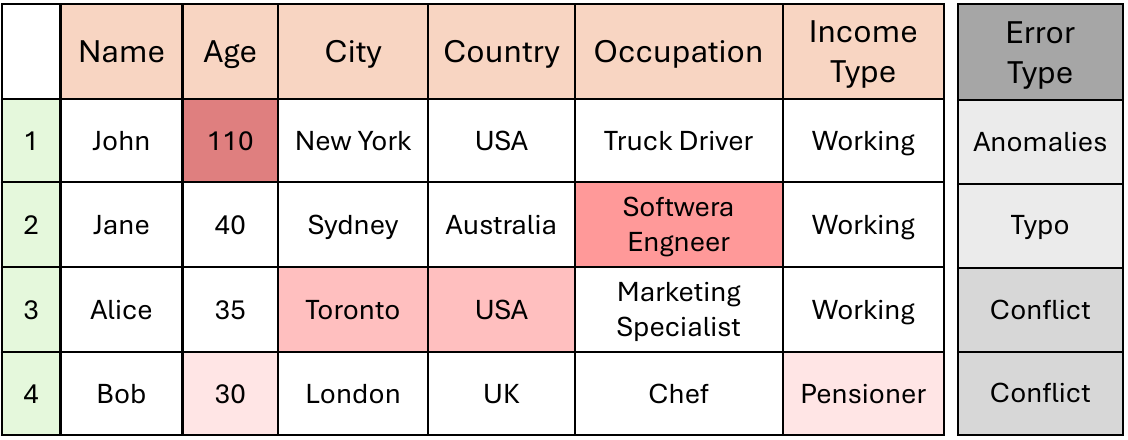}
\caption{Examples of Data Errors in tabular data: anomalies, typos, and conflicts between attributes.}
\label{fig:motivation}
\end{figure}

\noindent \textbf{Motivation:} 
To motivate our work, consider a dirty tabular dataset with four data quality errors that require verification. 
Figure~\ref{fig:motivation} shows common data errors: numeric anomalies, typos, and attribute mismatches. Tuple 1 includes a numeric anomalies value in the Age field, and tuple 2 has a typo in the Occupation field. 
Existing data quality validation systems, such as Amazon Deequ~\cite{schelter2018automating} and Google TFDV~\cite{caveness2020tensorflow}, can automatically generate constraints (e.g., the \textit{age} should be less than 100) to detect errors in tuples 1 and 2. 
However, these constraints are often difficult to set accurately, as valid cases may exist where age exceeds 100. In tuple 1, being employed as a truck driver at that age is highly unlikely and should be flagged as an error. Traditional systems often set such constraints either too strictly or too leniently, resulting in false positives or missed errors.
Tuple 3 contains a simple dependency issue, such as a city listed as \textit{Toronto} while the country is listed as the \textit{USA}. This type of error can often be addressed through expert-defined constraints. 
However, tuple 4 involves deeper dependencies, which is an illogical combination of age and income, referred to as hidden errors. Although data profiling~\cite{maydan1991efficient,naumann2014data,abedjan2017data} methods can discover such dependencies, existing automated data quality validation approaches only rely on expert-based strategies to identify such dependencies. 
As a result, they frequently overlook errors like those in tuple 4.


Our goal is to detect challenging errors, such as hidden errors (e.g., tuple 4), without relying on expert input, while also detecting ordinary errors like missing data and typos. Additionally, we provide suggestions for repairing the identified errors.

\noindent \textbf{Challenges:} 
Addressing the limitations of existing methods involves several challenges.
First, obtaining expert input for defining constraints is costly and often impractical, making an automated approach essential.
Second, existing methods struggle to detect subtle issues embedded within complex feature relationships, which we refer to as hidden data quality errors.
Third, the approach must be capable of identifying which specific features of a sample are contributing to the data quality issue, ensuring a precise assessment of problematic areas. 
Lastly, the model should provide suggestions to repair problematic features, while maintaining consistency with the overall clean data distribution.

\noindent \textbf{Our Solution:} To address these challenges, we propose DQuag, a multi-task learning framework, which integrates a GNN encoder~\cite{li2024graph} for feature embedding and dual decoders for data quality validation and repair. 
This combination leverages graph-based learning and multi-task optimization to detect and repair data quality issues.

The method includes the following key steps in the training phase: 
First, we construct a feature graph from clean tabular data to capture relationships between features. 
This graph representation enables the model to leverage relational dependencies inherent in the data. 
Next, a GNN encoder generates feature embeddings that encapsulate these relationships. 
The generated feature embeddings are then passed to two separate decoders. 
The Data Quality Validation Decoder reconstructs the original data from the feature embeddings, aiming to maximize the reconstruction accuracy for clean data and identify errors based on reconstruction loss. 
The Data Repair Decoder generates corrected values for problematic features. 
These decoders are optimized with distinct loss functions.

During the verification phase, we apply the trained model to unseen
datasets and use reconstruction errors to measure data quality, with
large errors indicating potential data quality issues. 
The repair decoder then suggests corrections for these features.
This approach detects both explicit and underlying data quality problems, reducing the need for expert-defined constraints and manual intervention. 

\noindent\textbf{Contributions:}  
This paper makes three key contributions. 
First, we introduce a novel advanced data quality validation and repair framework based on graph neural networks and multi-task learning, eliminating the need for expert-defined constraints. 
Second, we develop a novel dual-decoder architecture to address two critical tasks: data quality validation and data repair. 
The Data Quality Validation Decoder is designed to highlight potential quality issues, while the Data Repair Decoder provides correction suggestions.
Finally, we validate our approach through comprehensive and extensive experiments, thoroughly demonstrating its effectiveness and adaptability across various scenarios.

%% file: Sections/relatedwork.tex
\section{RELATED WORK}
\label{sec:relatedwork}
Research in data quality has traditionally emphasized the calculation of data quality dimensions and statistics to assess data quality. Fundamental components of this field include data profiling, which identifies issues like missing values and syntax violations~\cite{chiang2008discovering, ilyas2015trends, razniewski2011completeness}, and data quality assessment, which measures attributes such as accuracy, completeness, and consistency~\cite{DBLP:journals/jmis/WangS96, sidi2012data} to ensure datasets meet standards applicable to various domains~\cite{jayawardene2013curse, jayawardene2015analysis, YEGANEH201424}. 



Several studies have investigated data quality within ML pipelines, examining its influence on ML performance~\cite{Budach2022TheEO}, methods for high-quality data preparation~\cite{gupta2021data, Foroni2021EstimatingTE}, ensuring data completeness~\cite{karlavs2020nearest, schelter2020learning}, detecting data drift~\cite{tahmasbi2020driftsurf,concept-drift-96,dong2024efficiently}, and validating data quality~\cite{fadlallah2023context,sinthong2021dqdf,schelter2018automating,caveness2020tensorflow}. Frameworks like Dagger~\cite{rezig2020dagger} and Mlinspect~\cite{grafberger2021mlinspect} incorporate advanced debugging and interactive query mechanisms to streamline systematic data management. Furthermore, tools such as HoloClean~\cite{rekatsinas2017holoclean} and ActiveClean~\cite{krishnan2016activeclean} extend these efforts to data cleaning and repair, leveraging probabilistic models and active learning for iterative dataset refinement.

Existing quality validation systems typically use statistical methods or statistics to train machine learning models for detecting data quality problems. 
These methods often rely on predefined rules and constraints that require expert adjustment~\cite {fadlallah2023context, sinthong2021dqdf}, e.g., automated data quality verification systems, such as Deequ~\cite{schelter2018automating} and TFDV~\cite{caveness2020tensorflow}, which generate constraints automatically. 
However, these constraints are not always accurate and often need expert fine-tuning to effectively detect hidden data relationships and dependencies~\cite{schelter2019differential}. 
Other systems like ADQV~\cite{redyuk2021automating} use k-nearest neighbors to evaluate data quality by computing limited metrics, but fail to detect hidden errors and cannot pinpoint the incorrect samples.

Our approach uses a multi-task learning framework that combines GNN encoder~\cite{li2024graph} and dural decoders. 
While GNNs and other machine learning models are used for data imputation~\cite{telyatnikov2023egg, cappuzzo2024relational, spinelli2020missing}, anomaly detection~\cite{du2022graph}, and fraud detection~\cite{rao2020xfraud, liu2021pick}, their use in an integrated manner for direct data quality validation remains underexplored.
Our work applies GNN to automatically validate data quality, focusing on both explicit data errors and complex interdependencies without relying on expert-defined rules and manual adjustments and optimizes for both detection and repair tasks, setting our approach apart from conventional data management techniques that typically rely on isolated, task-specific methods.

%% file: Sections/systems.tex
\section{OUR APPROACH: DQuaG}

\begin{figure*}[tb]
\centering
\includegraphics[width=0.7\textwidth]{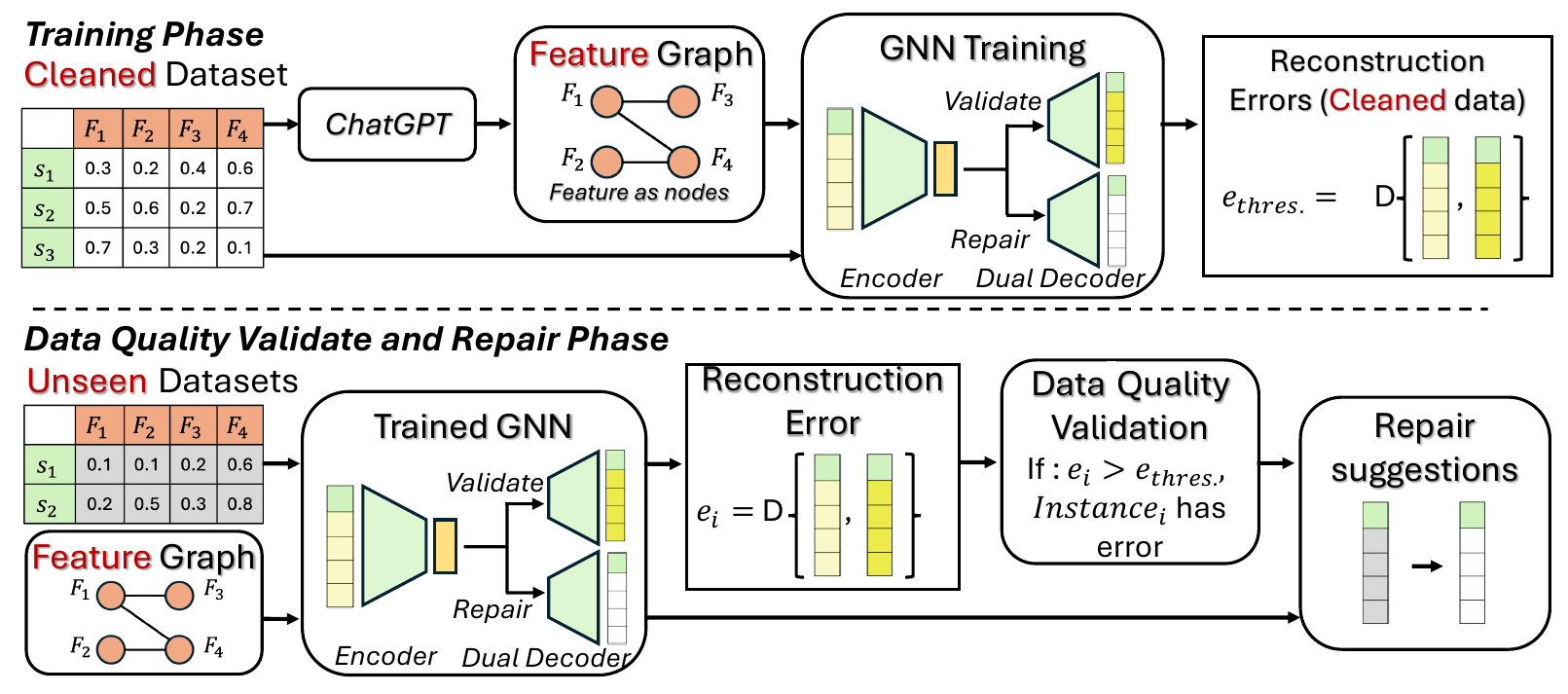}
\caption{Data Quality Validation Framework Using GNN. Top: Training on clean data. Bottom: Validating unseen data by reconstruction error comparison.}
\label{fig:framework}
\end{figure*}


In this section, we present DQuaG (Data Quality Graph), a novel approach for data quality validation. 
Figure~\ref{fig:framework} illustrates the framework of our approach, which includes two main phases: model training on a clean dataset and data quality validation and repair for new data. 
In Phase 1, we train a model using a clean dataset to learn the normal patterns and relationships between features. 
In Phase 2, we use the trained model to assess the quality of new data and provide repair suggestions for any detected errors. 

\subsection{Phase 1: Training GNN on Clean Data}
We assume the availability of a high-quality, clean dataset, $\mathcal{D}_{\text{clean}}$, that has undergone rigorous quality control and is free from errors. This dataset serves as the foundation for training our model. 
For feature encoding and normalization, categorical features are converted to numerical form using label encoding, where the encoder is fitted on both clean data and any possible future data to ensure consistency. 
For numerical features, we apply min-max normalization to scale values to the range [0, 1], which helps improve training stability and ensures that all features are on a comparable scale.

\subsubsection{\textbf{Feature Graph Construction}}

We use ChatGPT-4~\cite{openai2024gpt4} to automate the feature graph construction. 
Given a clean dataset, we extract the feature names ($F$) and their descriptions ($D$) from the data source. We then randomly sample 100 data points from the dataset, denoted as ($S$). 
These feature names, descriptions, and sample points are provided to ChatGPT-4 in a structured format to infer potential relationships between features. 
The output from ChatGPT-4 is a JSON file capturing feature relationships, which we denote as  \(\text{Feature\_Relationships} = \{ (f_i, f_j) \mid f_i, f_j \in F \}\), indicating that there is a relationship between features \( f_i \) and \( f_j \).

\begin{tcolorbox}[
    sharp corners=south,
    colback=white!98!black,
    colframe=white!45!black,
    boxrule=0.5mm,
    width=0.48\textwidth,
    enlarge left by=0mm,
    enlarge right by=0mm,
    arc=5mm,
    outer arc=3mm,
    fonttitle=\bfseries,
    title=Prompt for Feature Relationship Inference,
    before upper=\par\small,
    after upper=\par\small
]
\vspace*{-0.2cm}
\small
Given the following information, please infer the relationships between features. Provide your output in JSON format, capturing the type of relationships.\newline
\textbf{Feature Names:} {List of feature names ($F$)}\newline
\textbf{Feature Descriptions:} {List of descriptions ($D$) for each feature}\newline
\textbf{Sample Data Points:} {100 data samples ($S$) from the dataset}\newline
\textbf{Output:} Please return a JSON object in the format:
\begin{verbatim}
{"relationships": [{"feature1", "feature2"}, 
                   {"feature3", "feature4"}, ...]}
\end{verbatim}
\vspace*{-0.3cm}
\end{tcolorbox}

Using these relationships, we construct the knowledge-based feature graph \( G = (V, E) \), where \( V \) represents features and \( E \) represents edges indicating relationships between features.

\subsubsection{\textbf{GNN Model Architecture}}
Our model architecture combines the strengths of different graph neural network variants to effectively capture complex feature relationships. 
The architecture consists of three main components: an improved GNN encoder that fuses Graph Attention Network (GAT)~\cite{velivckovic2017graph} layers and Graph Isomorphism Network (GIN)~\cite{xu2018powerful} layers, and two specialized decoders for quality validation and repair suggestion generation.

\noindent{\textbf{GNN Encoder (GAT + GIN)}.
Our encoder consists of four layers: alternating \textit{Graph Attention Network (GAT)} and \textit{Graph Isomorphism Network (GIN)} layers, in the order of GAT-GIN-GAT-GIN. 
This design is inspired by recent findings in the field of graph representation learning that demonstrate how combining different types of graph layers can yield improved performance in feature extraction and relational representation tasks~\cite{zhang2019heterogeneous}. 
Our experimental results demonstrate the advantages of this structure.

The GAT layers compute attention weights between connected features, enabling the model to adaptively assign importance to significant relationships in the data. This allows the model to focus on critical connections and ignore irrelevant information, which enhances its ability to learn meaningful feature representations. Our approach uses GAT layers, which automatically learn edge weights through attention mechanisms during training. This eliminates the need to manually assign weights in the initial feature graph. 

The GIN layers aggregate feature information from neighboring nodes to capture structural information more effectively. By using GIN, the encoder gains a strong ability to represent the underlying structure of the data, preserving key relationships crucial for data quality validation and repair tasks.

This alternating GAT and GIN structure enhances the model's ability to both prioritize important features and learn intricate structural relationships, thereby making it more effective at representing complex feature dependencies in the data.
Specifically, the GNN encoder processes the feature graph $G = (V, E)$ along with the input data matrix $\mathbf{X} \in \mathbb{R}^{n \times d}$, where $n$ is the number of nodes (features) and $d$ is the dimensionality of each feature vector. The output from the GNN encoder is a feature embedding matrix $\mathbf{Z} \in \mathbb{R}^{n \times h}$, where $h$ represents the size of the learned feature embeddings.

\noindent{\textbf{Dual Decoder Structure}.}  
Our model employs two separate decoders to address the tasks of Data Quality Validation and Repair Suggestion, enabling focused optimization for each objective.

\textit{Data Quality Validation Decoder} is responsible for reconstructing the original feature space from the learned embeddings, denoted as \(\mathbf{Z}\). 
The primary objective of this decoder is to learn the correct patterns from clean data and reconstruct the features in a way that captures the underlying structure of the dataset. This allows us to identify abnormalities by measuring reconstruction errors. We have designed a unique loss function that ensures the model focuses on learning accurate representations of clean data while effectively distinguishing abnormal samples.

For normal data samples, the decoder should ideally have a low reconstruction error, as the learned embeddings should effectively capture the true relationships between the features, resulting in an accurate reconstruction. For abnormal samples, the reconstruction error will be higher, indicating that these samples do not conform to the learned patterns from the clean data.
The reconstruction loss is defined as 
\( L_{\text{validation}} = \frac{1}{N} \sum_{i=1}^{N} w_i \left\| \mathbf{X}_i - \hat{\mathbf{X}}_i \right\|_2^2 \),
where \(\mathbf{X}_i\) represents the original input features, and \(\hat{\mathbf{X}}_i\) is the reconstructed feature vector for the \(i\)-th sample. The weights \(w_i\) are assigned to each sample based on its reconstruction error.

We assign larger weights to normal data samples (\(w_i\) s higher for samples with smaller reconstruction errors), giving them a greater influence in minimizing their reconstruction loss. This encourages the model to accurately reconstruct the normal data and effectively learn the correct data distribution. 
For samples with potential quality issues, the weights \(w_i\) are reduced, meaning that their influence on the overall loss is diminished. 
This allows the model to focus on minimizing the reconstruction loss for normal data while maintaining high reconstruction errors for problematic samples during the backpropagation. 
By using this weighting mechanism, we ensure that the validation decoder can distinguish between normal data and data with potential issues based on reconstruction errors.

\textit{Data Repair Decoder}, on the other hand, takes the same learned embeddings \(\mathbf{Z}\) as input, but its goal is different: it aims to suggest repaired values for features identified as erroneous. 
Unlike the Data Quality Validation Decoder, which reconstructs data to highlight discrepancies, the Data Repair Decoder attempts to produce an output that aligns with the clean, underlying data distribution, effectively suggesting corrections for the detected errors. 
The objective of this decoder is defined through the following loss function:
\(
L_{\text{repair}} = \frac{1}{N} \sum_{i=1}^{N} \left\| {\mathbf{X}}_i - \tilde{\mathbf{X}}_i \right\|_2^2
\).
Here, \(\tilde{\mathbf{X}}_i\) represents the feature values repaired by the decoder, while \(\mathbf{X}_i\) stands for the corresponding clean feature values from the input dataset. Since the input is already clean, \(\mathbf{X}_i\) can directly serve as the target for the repair task.

The combination of these two decoders is essential for effectively handling data quality issues. 
The overall loss function is a weighted sum of the validation and repair losses:
\(
L_{\text{total}} = \alpha L_{\text{validation}} + \beta L_{\text{repair}}
\),
where \(\alpha\) and \(\beta\) are hyperparameters used to balance the contributions of reconstruction and repair, both of which are set to 1 in our experiments.

The two decoders serve different purposes: the \textit{Data Quality Validation Decoder} is optimized to detect data issues by maximizing reconstruction errors for problematic instances, while the \textit{Data Repair Decoder} aims to provide realistic corrections for identified issues. By separating these tasks, the model avoids conflicting optimization goals, ensuring it is both effective at identifying problems and providing reliable repairs.

\noindent{\textbf{Multi-Task Learning Framework}.}
The encoder is shared between the quality validation and repair tasks, while each task-specific decoder learns independently. This multi-task framework enables the model to exploit shared information between these tasks, allowing the model to learn a unified representation beneficial for both.

\subsubsection{\textbf{Training Process.}}
We train the model on the clean dataset using an optimizer Adam to minimize $L_{\text{total}}$. 

\subsubsection{\textbf{Collecting the statistics of reconstruction errors.}}
During training, we record the reconstruction error for each instance.
The reconstruction error is essentially the loss for each instance.
Let $e_i$ denote the reconstruction error for instance $i$, and let $\mathcal{E}$ be the set of all reconstruction errors from the clean dataset. 
Given that even cleaned datasets may contain undetected errors, we do not set the maximum reconstruction error as the threshold for identifying problematic instances. 
Instead, we set the threshold at the 95th percentile of \(\mathcal{E}\), denoted as \( e_{threshold} \).

Instances in the next phase with reconstruction errors above \( e_{threshold} \) are flagged as potentially problematic.

\subsection{Phase 2: Data Quality Validate and Repair}

\subsubsection{\textbf{Data Quality Validation Process}}
In this phase, we validate the quality of incoming data by comparing it to the patterns learned during model training.

The new unseen data is preprocessed in the same manner as the clean dataset to ensure consistency in feature encoding, normalization, and feature graph construction. These new unseen datasets must keep the same schema as the original clean dataset.

\noindent{\textbf{Detecting Data Quality Issues by Reconstruction Errors}.}
After preprocessing, the model uses the validation decoder to reconstruct the features of the new data. 
For each data instance, the reconstruction error $e_i$ is calculated. We then obtain a list of reconstruction errors, denoted as \(\mathcal{E}_{\text{new}}\).
Next, we compare each reconstruction error in \(\mathcal{E}_{\text{new}}\) with the threshold \( e_{threshold} \) from the clean dataset. 
We calculate the proportion of instances in the new dataset with reconstruction errors exceeding \( e_{threshold} \), denoted as \( R_{error} \). 
Since the threshold was set at the 95th percentile for the clean dataset, we expect around 5\% of clean data instances to exceed this value. 

To account for data variability, if \( R_{error} \) exceeds \( 5\% \times n \), we classify the new dataset as problematic. This means if more than \( 5n\% \) of instances in the new dataset have errors greater than \( e_{threshold} \), we will report the dataset has data quality issues. The parameter \( n \) can be adjusted based on observed reconstruction errors after deployment.
In our experiments, we set \( n = 1.2 \), which exhibited good performance.
Finally, we report the indices of all instances in the new dataset with reconstruction errors above \( e_{threshold} \), clearly identifying problematic samples.

\noindent{\textbf{Detecting Feature Errors}.}
Each instance's reconstruction error \( e \) is a list corresponding to each feature's loss. To identify specific problematic features, we detect outliers with significantly higher reconstruction errors.
For an instance \( \mathbf{x}_i \), let \( \mathbf{e}_i = [e_{i1}, e_{i2}, \ldots, e_{in}] \) be the reconstruction errors for the \( n \) features. We calculate the mean \( \mu_i \) and standard deviation \( \sigma_i \) of the errors. Features with errors greater than \( \mu_i + 5\sigma_i \) are flagged as problematic.


\subsubsection{\textbf{Repair Suggestion Generation}}
In this phase, we provide repair suggestions for detected errors to improve the quality of the data for downstream use.
The repair decoder is used to generate a repaired feature vector, which includes suggested repaired values for all features. 
In the previous step, we flagged which specific instances and features were problematic. 
Then we selectively apply modifications only to the flagged problematic features. For categorical features, the repair decoder predicts the most likely corrected category, while for numerical features, it predicts a value that aligns with the learned data distribution. 

%% file: Sections/evaluation.tex
\section{EXPERIMENTAL EVALUATION}

\subsection{Experimental Setup}
The experiments were conducted using Python 3.11 and PyTorch 1.12.1. All computations were performed on an NVIDIA A100 GPU. The source code and data have been made available\textsuperscript{\ref{footnote:code}}.

\footnotetext[1]{\label{footnote:code}Source code and data: \url{https://github.com/SiSijie/DQuaG}}

\subsubsection{Datasets.} We evaluate the robustness and generality of our approach using datasets with varied error types and data structures, following methodologies outlined in prior research~\cite{redyuk2021automating}.

\noindent \textit{Datasets with ground-truth errors.}
\textbf{Airbnb Data\cite{airbnb}} which contains information about Airbnb listings in New York City, including attributes such as price, location, and property type; and
\textbf{Chicago Divvy Bicycle Sharing Data\cite{divvy}} includes trip data from the Divvy bike-sharing program in Chicago, with details like the trip duration, start and end locations, and bike ID. 
\textbf{Google Play Store Apps Data\cite{google_play}} includes app data, capturing ratings, downloads, categories, and trends in app performance.
For these datasets, we have uncleaned versions with real-world errors.
We involved data-cleaning techniques to create cleaned version datasets, such as removing duplicates, handling missing data, and filtering out illogical records.


\noindent \textit{Datasets without ground-truth errors.} 
\textbf{New York Taxi Trip Data\cite{nyc_taxi}} which comprises taxi trip records in New York City, detailing pickup and dropoff locations, fares, and trip durations;
\textbf{Hotel Booking Data\cite{antonio2019hotel}} contains booking information for a city hotel and a resort hotel; and
\textbf{Credit Card Data\cite{credit_card}} includes information on credit card applications.
For these datasets, we directly utilized clean versions of data from reliable sources~\cite{credit_card,antonio2019hotel,nyc_taxi}. These datasets were publicly released, carefully collected, and cleaned before use. From these clean versions, we generated four types of data errors to create the corresponding dirty version datasets.

Clean datasets are not entirely error-free. Clean data is defined as having higher quality than dirty datasets and meeting the user's standards.

\vspace*{-0.2cm}
\subsubsection{Synthetic Errors in Datasets without ground-truth errors.}

We simulate three ordinary errors and propose two potential hidden errors to evaluate the effectiveness of our approach:


\textit{Ordinary Errors:} We introduce three types of errors affecting 20\% of values in three selected attributes: missing Values arise from empty cells due to collection or integration errors; numeric anomalies occur when sensor malfunctions or scaling issues result in out-of-range values; and String typos are caused by spelling errors simulated by randomly replacing letters with neighboring keys on a "qwerty" keyboard.


\textit{Hidden Errors:}
\textbf{Logical and Temporal Conflicts between Attributes}, which occur when related attributes contain values that are either conflicting or illogical when considered together, or when time-related data does not follow chronological logic. 
In the Credit Card dataset, we set two hidden conflicts. 
First, a conflict arises when \textit{DAYS\_EMPLOYED} exceeds \textit{DAYS\_BIRTH}, implying employment before birth. 
Second, a conflict involves \textit{AMT\_INCOME\_TOTAL},
\textit{NAME\_EDUCATION\_TYPE}, and \textit{OCCUPATION\_TYPE}, producing improbable combinations (e.g., high education and advanced occupation but extremely low income) that simple domain rules rarely cover.
In the Hotel Booking dataset, a hidden error was generated for bookings labeled \textit{customer\_type} as 'Group' with zero \textit{adults} and more than zero \textit{babies}, conflicting with logic.


\subsubsection{Baselines}
We selected four SOTA baselines, each widely recognized for their effectiveness in detecting and validating data quality issues.
\noindent\textbf{Deequ~\cite{schelter2018automating}:} A tool by Amazon for scalable data quality validation, using constraints and metrics. \textit{Deequ auto} generates constraints automatically, while \textit{Deequ expert} incorporates expert-tuned adjustments for higher precision.
\noindent\textbf{TFDV~\cite{caveness2020tensorflow}:} TensorFlow Data Validation for scalable ML pipeline validation, automatically detecting anomalies and schema violations. \textit{TFDV auto} uses auto-generated constraints, while \textit{TFDV expert} supports expert fine-tuning.
\noindent\textbf{ADQV~\cite{redyuk2021automating}:} A tool leveraging adaptive learning to dynamically adjust validation criteria, excelling in evolving datasets.
\noindent\textbf{Gate~\cite{shankar2023automatic}:} A machine learning-based method for automated detection and correction of data quality issues, offering an adaptive alternative to rule-based systems.

Similar to previous work~\cite{redyuk2021automating}, we manually performed the fine-tuning work required by experts for Deequ and TFDV.


\subsection{Accuracy of Synthetic Error Detection}
\begin{table}[tb]
\centering
\caption{Accuracy and recall across different methods and two datasets with synthetic data errors (Ordinary errors: N = Numeric Anomalies, S = String Typos, M = Missing Values; Hidden Errors: Conflicts = Logical and Temporal Conflicts between Attributes). \small{Note: * Indicates average value.}}
\label{table:consolidated_validation}
\footnotesize 
\begin{tabular}{lllcc}
\hline
\textbf{Dataset} & \textbf{Error Types} & \textbf{Methods} & \textbf{Acc.} & \textbf{Recall} \\
\hline
\multirow{5}{*}{Hotel Booking} 
 & N, S, M & Deequ auto & 0.530* & 1 \\
 & N, S, M & Deequ expert & 1 & 1 \\
 & N, S, M & TFDV auto, expert & 1 & 1 \\
 & N, S, M & ADQV & 0.963* & 1 \\
 & N & Gate & 0.500 & 0 \\
  & S, M & Gate & 0.980* & 0.960* \\
 & N, S, M & \textbf{DQuaG} & 1  & 1 \\
\cline{2-5}
 & Conflicts & Deequ expert & 0.500 & 0 \\
 & Conflicts & TFDV expert & 0.500 & 0 \\
& Conflicts &  ADQV & 0.970 & 1 \\
& Conflicts &  Gate & 0.820 & 0.640 \\
 & Conflicts & \textbf{DQuaG} &  \textbf{1} & \textbf{1} \\
\hline
\multirow{6}{*}{Credit Card} 
 & N, S, M & Deequ auto & 0.550* & 1 \\
 & N, S, M & Deequ expert & 0.970* & 1 \\
 & S, M & TFDV auto & 1 & 1 \\
 & N  & TFDV auto &  0.500 & 0 \\
 & N, S, M & TFDV expert &  1 & 1 \\
 & N, S, M  & ADQV & 0.960* & 1 \\
 & N, S, M  & Gate & 0.510* & 1 \\
 & N, S, M  & \textbf{DQuaG} & 1 & 1 \\
 \cline{2-5}
 & Conflicts-1 & Deequ expert & 0.500 & 0 \\
& Conflicts-1 & TFDV expert & 0.500 & 0 \\
& Conflicts-1 & ADQV & 0.500 & 1 \\
& Conflicts-1 & Gate & 0.510 & 1 \\
 & Conflicts-1 & \textbf{DQuaG} & \textbf{1} & \textbf{1}\\
  \cline{2-5}
 & Conflicts-2 & Deequ expert & 0.500 & 0 \\
 & Conflicts-2 & TFDV expert & 0.500 & 0 \\
 & Conflicts-2 & ADQV & 0.960 & 1 \\
 & Conflicts-2 & Gate & 0.560 & 1 \\
 & Conflicts-2 & \textbf{DQuaG} & \textbf{1} & \textbf{1} \\ 
\hline
\end{tabular}%

\end{table}

We used a clean dataset, randomly sampling 10\% to generate 50 batches of clean data, and did the same with a dirty dataset to generate 50 batches of dirty data. We then used these 100 batches to test
our method and baselines.

According to Table~\ref{table:consolidated_validation}, our method performs well in detecting both ordinary and hidden errors, achieving accuracy and recall of 1. 
For ordinary errors, Deequ-auto and TFDV-auto methods are inaccurate due to overly strict or soft constraints but perform well after expert tuning. However, they cannot detect hidden errors. 
ADQV can automatically detect ordinary errors but fails to identify hidden errors. 
For hidden errors in the Hotel Booking dataset, ADQV shows an accuracy of 0.97 but actually flags change in numeric feature distribution, missing the real issues. 
For the Gate, the results are also unstable. The constraints it sets are too strict and cannot distinguish between dirty and clean datasets. It also cannot distinguish well between hidden errors.
The experimental results highlight our method's robustness, particularly in identifying complex data interdependencies.

\subsection{Accuracy of Real-World Error Detection}
\begin{figure}[tb]
\centering
\includegraphics[width=1.05\linewidth]{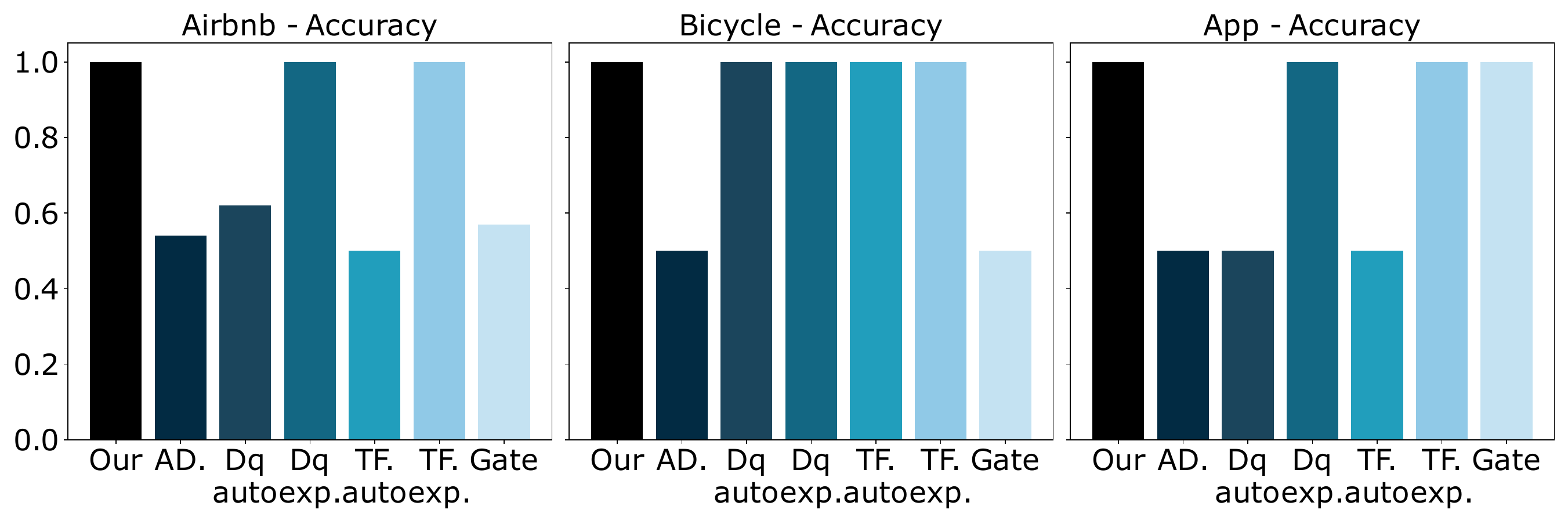}
\caption{Accuracy across different methods and two datasets with real-world data errors. (All methods have Recall=1)}
\label{fig:accuracy}
\end{figure}

We used the same method to generate 100 test batches from Airbnb and Bicycle datasets as in Section 4.2. In Figure~\ref{fig:accuracy}, our approach showed excellent performance in detecting problematic data in both real-world datasets, achieving an accuracy of 1. 
ADQV and Gate performed poorly, flagging all batches due to overly strict error detection. 
Deequ auto and TFDV auto performed poorly on the Airbnb dataset but well on the bicycle dataset. Deequ expert and TFDV expert performed well on both datasets but required manual tuning of constraints. 
Therefore, our method effectively detects real-world errors without the need for manual intervention.

\subsection{Comparison with Encoder Architectures}
\label{sec:gnn_comparison}
To validate the effectiveness of the GAT+GIN architecture, 
we tested five encoder architectures: \textit{Graph2Vec}~\cite{narayanan2017graph2vec}, \textit{GCN}, \textit{GCN+GAT}, \textit{GCN+GIN}, and \textit{GAT+GIN} on Airbnb and Bicycle datasets. Table~\ref{tab:architectures} shows that \textit{GAT+GIN} achieves the highest difference in flagged errors, reflecting its stronger ability to distinguish clean from dirty data. 
We attribute this to GAT’s attention mechanism (focusing on important neighbors) and GIN’s injective aggregation (capturing nuanced relationships). 
For hyperparameters, all models used four layers, a hidden dimension of 64, a learning rate of 0.01, and a batch size of 128.

\begin{table}[tb]
\centering
\footnotesize
\caption{Difference (\%) in flagged errors for clean vs.~dirty data. 
(Higher is better.)}

\begin{tabular}{lccccc}
\toprule
\textbf{Dataset} & \textbf{Graph2Vec} & \textbf{GCN} & \textbf{GCN+GAT} & \textbf{GCN+GIN} & \textbf{GAT+GIN} \\
\midrule
Airbnb  & 2.72 & 1.83 & 2.60 & \textbf{4.55} & \textbf{4.17} \\
Bicycle & 21.49 & 11.06 & 12.36 & 17.51 & \textbf{21.72} \\
\bottomrule
\end{tabular}

\label{tab:architectures}
\end{table}

\subsection{Scalability Analysis}
\begin{figure}[tb]
\centering\includegraphics[width=0.7\columnwidth]{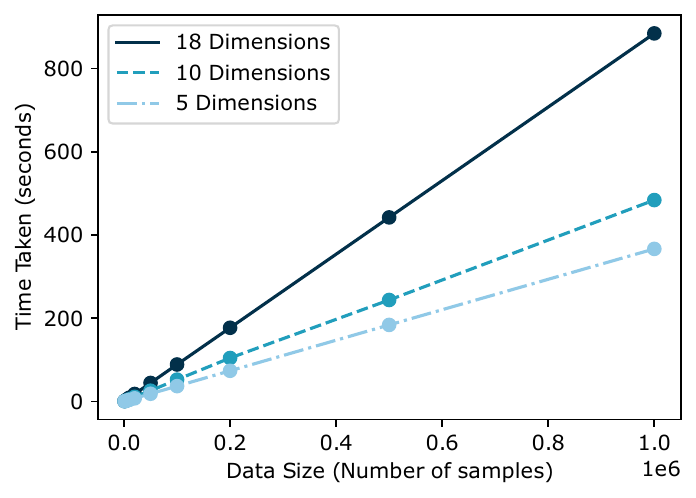}
\caption{Scalability Analysis: data quality validation time of our approach, when varying the data dimensionality and data size, on the New York Taxi dataset.}
\label{fig:Scalability_log}
\end{figure}

Figure~\ref{fig:Scalability_log} illustrates the scalability analysis of our data quality validation method on the New York Taxi dataset, showing the validation time across different data dimensions (5, 10, and 18) and varying data sizes (up to 1 million samples).
The results demonstrate that as the dataset size and the number of dimensions increases, our method's computation time increases linearly rather than exponentially, which is logical. When processing datasets with millions of data points, our method only takes ten minutes. These findings indicate that our method is scalable.

To further assess scalability and robustness, we conducted an experiment using different sample sizes (10 to 1000). The accuracy results are summarized in Table~\ref{tab:accuracy_summary}. The sample size represents the number of new data instances used for quality validation.
        
\begin{table}[ht]
    \centering
    \small
    \caption{Summary of Overall Accuracy for Different Sample Sizes across 3 Datasets.
    }
    \begin{tabular}{@{}lcccccc@{}}
        \toprule
        \textbf{Sample Size} & \textbf{10} & \textbf{20} & \textbf{50} & \textbf{100} & \textbf{500} & \textbf{1000} \\ \midrule
        \textbf{Airbnb (Accuracy \%)}  & 85.0 & 93.0 & 99.0 & 99.0 & \textbf{100.0} & \textbf{100.0} \\
        \textbf{Bicycle (Accuracy \%)} & 86.0 & 92.0 & 89.0 & 97.0 & \textbf{100.0} & \textbf{100.0}
        \\
        \textbf{NY Taxi (Accuracy \%)} & 83.0 & 89.0 & 98.0 & 97.0 & \textbf{100.0} & \textbf{100.0}
        \\ \bottomrule
    \end{tabular}
    \label{tab:accuracy_summary}
\end{table}

The experimental results show that as the sample size increases, the model's accuracy also improves. When the sample size exceeds 500, the model achieves 100\% accuracy across both datasets, indicating stable performance at larger scales. These findings highlight the effectiveness of our method for large sample sizes but also suggest limitations in scenarios with smaller data availability.

\subsection{Data Repair Evaluation}
To assess the effectiveness of our data repair process, we conducted experiments on the Airbnb and Bicycle datasets. 
According to our data quality validation results of the Airbnb dataset, The error rate of the original dirty dataset was 10.52\%. 
After applying repair suggestions generated by our repair decoder, the error rate was reduced to 4.97\%, closely matching the error rate of 4.95\% observed in the clean dataset. 
For the results of the Bicycle dataset, the repair decoder reduced the error rate of the dirty data from 21.11\% to 2.75\%. 
Importantly, the repaired dataset was classified as clean under our quality validation framework.

%% file: Sections/conclusion.tex
\section{Conclusions and Future Work}
We proposed DQuaG, a novel multi-task learning framework that combines a GNN encoder with dual decoders for data quality validation and repair. 
Our approach effectively detects and repairs data quality issues without expert-defined constraints or manual interventions. By utilizing a dual-decoder structure, our model independently optimizes data quality validation and repair tasks, with a specially designed validation decoder loss that enhances the detection of erroneous data. 
We evaluated our approach using real-world datasets containing inherent data quality issues, where it performed well. However, in extreme scenarios—such as datasets with only a few thousand samples where only a single erroneous instance exists—our method may lack sensitivity. In such cases, traditional statistical methods often prove to be more accurate, as they can better identify isolated anomalies within smaller datasets. 
We aim to further refine our approach, improving its robustness in extreme conditions.
Additionally, we plan to extend our approach to encompass post-validation tasks, such as data cleaning and data selection. 
To improve the interpretability of our models, we will focus on optimizing the GNN frameworks.




